\documentclass[prd,twocolumn,nofootinbib]{revtex4}
\usepackage{amsmath,amssymb,bm,graphicx}

\begin{document}

\title{Non-parametric reconstruction of an inflaton potential from Einstein-Cartan-Sciama-Kibble gravity with particle production}

\author{Shantanu Desai$^{1,2}$}
\altaffiliation{shantanu@usm.lmu.de}

\author{Nikodem J. Pop{\l}awski$^{3}$}
\altaffiliation{NPoplawski@newhaven.edu}

\affiliation{$^{1}$Faculty of Physics, Ludwig-Maximilians University, Scheinerstrasse 1, 81679 Munich, Germany}
\affiliation{$^{2}$Excellence Cluster Universe, Boltzmannstrasse 2, 85748 Garching, Germany}
\affiliation{$^{3}$Department of Mathematics and Physics, University of New Haven, 300 Boston Post Road, West Haven, CT 06516, USA}

\begin{abstract}
The coupling between spin and torsion in the Einstein-Cartan-Sciama-Kibble theory of gravity generates gravitational repulsion at very high densities, which prevents a singularity in a black hole and may create there a new universe.
We show that quantum particle production in such a universe near the last bounce, which represents the Big Bang gives the dynamics that solves the horizon, flatness, and homogeneity problems in cosmology.
For a particular range of the particle production coefficient, we obtain a nearly constant Hubble parameter that gives an exponential expansion of the universe with more than 60 $e$-folds, which lasts about $\sim 10^{-42}$ s.
This scenario can thus explain cosmic inflation without requiring a fundamental scalar field and reheating.
From the obtained time dependence of the scale factor, we follow the prescription of Ellis and Madsen to reconstruct in a non-parametric way a scalar field potential which gives the same dynamics of the early universe.
This potential gives the slow-roll parameters of cosmic inflation, from which we calculate the tensor-to-scalar ratio, the scalar spectral index of density perturbations, and its running as functions of the production coefficient.
We find that these quantities do not significantly depend on the scale factor at the Big Bounce.
Our predictions for these quantities are consistent with the Planck 2015 observations.
\end{abstract}



\maketitle


\section{Introduction}
It has been known since the 1970s that the standard hot Big Bang model suffers from the horizon, flatness, and homogeneity problems~\cite{Dicke} and there must be another dynamical mechanism prior to Big Bang nucleosynthesis to alleviate these problems.
Currently, the most widely accepted solution to these problems is the process of cosmic inflation, which is a brief period of exponential expansion, where the Universe is temporarily in a de Sitter phase dominated by the vacuum energy~\cite{Starobinsky,Kazanas,Sato,Guth,Linde,Albrecht}.
When the inflationary epoch ends, the Universe starts decelerating, which is followed by the phenomenon of reheating when the energy in the inflaton field is dumped into standard model particles.
Most of the generic models of cosmic inflation are usually due to scalar field in the slow-roll approximation, where potential energy of the field dominates over its kinetic energy. 
In the last three decades a plethora of  inflationary potentials have been constructed, and currently single-field models are observationally the most favored ones~\cite{Martin}.
The basic predictions of single scalar field slow-roll inflation models, such as flatness, super-horizon correlations, adiabatic density perturbations, nearly scale-invariant spectrum of curvature perturbations, and no observable non-gaussianity, have been verified by the Cosmic Microwave Background (CMB) observations from Planck and WMAP~\cite{Planck2013,Planck2015,WMAP}.
Inflation also provides a mechanism to seed the density perturbations which give rise to the observed structure in the universe.
However despite these predictions, many concerns have been raised about conceptual problems with models of inflation based on single-field scalar potentials~\cite{Ijjas}.
Moreover, most models of inflation do not address other problems such as Big Bang singularity, origin of the matter-antimatter asymmetry, arrow of time, or the nature of dark matter/dark energy.

In this paper, we consider a model of cosmic inflation which is motivated by an extension of general relativity, called the Einstein-Cartan-Sciama-Kibble (ECSK) theory of gravity~\cite{ScKi1,ScKi2,review,Niko}.
The ECSK theory naturally extends the metric general relativity (GR) by removing its constraint that the affine connection is symmetric and thus allowing the exchange between the orbital and spin components of the total angular momentum \cite{req}.
The antisymmetric part of the connection, which is the torsion tensor, becomes a dynamical variable related to the spin density of matter.
The source of torsion is provided by the spin of fermions approximated as a spin fluid \cite{Niko,HHK,Gas}.
The ECSK theory may also solve the problem of divergent integrals in quantum field theory by providing fermions with spatial extension and thus introducing an effective ultraviolet cutoff for their propagators \cite{Dir}.
At extremely high densities existing in black holes and in the very early Universe, the minimal spin-torsion coupling manifests itself as a repulsive force, which avoids the formation of singularities from fermionic matter \cite{HHK,Gas,Dir,avert1,avert2,Kuc,torsion1,torsion2,torsion3,Alexander,Magueijo}.
Accordingly, the singular Big Bang is replaced by a non-singular Big Bounce, before which the Universe was contracting \cite{torsion1,Kuc}.
The ECSK theory agrees with all solar system, binary pulsar and cosmological tests of GR, since even at nuclear densities, the contribution from torsion to the Einstein equations is negligible.
Torsion therefore provides the simplest and most natural mechanism that solves the singularity problem of the standard Big Bang cosmology.

In this paper, we study how a scenario of a closed universe inside a black hole with torsion and quantum particle production \cite{Par,Zel,ZeSt}, which was introduced in \cite{torsion2014}, can provide a viable model of inflation.
We numerically solve the Friedmann equations which describe the dynamics of such a universe~\cite{torsion2014} and calculate the number of $e$-folds as a function of the particle production coefficient.
In order to make predictions for observables such as the tensor-to-scalar ratio and scalar spectral index of density fluctuations, we reconstruct a dynamically equivalent single-field scalar potential from the time dependence of the scale factor calculated in our model.
We then calculate the slow roll parameters from this non-parametric potential by applying the quantization of a scalar field in curved spacetime \cite{Mukhanov}, and compare our results to the Planck 2015 results \cite{Planck2015}.
We emphasize that our model of inflation does not contain a fundamental scalar, and the reconstruction of an inflation potential is only a mathematical technique we use to calculate the scalar spectral index and tensor-to-scalar ratio. We also briefly discuss some other models of inflation based upon torsion and compare them with our approach.


\section{Universe in a black hole with torsion}
A model of a closed universe in a black hole with torsion was introduced in~\cite{torsion2014}.
In this section, we briefly review the ECSK theory of gravity which provides torsion as a mechanism for this scenario and then write the equations describing the dynamics of such a universe.
We use the same notation as \cite{torsion2014}.

In the ECSK theory, the antisymmetric part of the affine connection (the torsion tensor) does not vanish but is determined by the field equations which are obtained by varying the total action for the gravitational field and matter with respect to the metric and torsion tensors~\cite{ScKi1,ScKi2,review,Niko}.
Varying the action with respect to the torsion gives the Cartan equations that relate algebraically the torsion of spacetime to the canonical spin tensor of matter.
Varying the action with respect to the metric gives the Einstein equations that relate the curvature of spacetime to the canonical energy-momentum tensor of matter.
The Einstein and Cartan equations can be combined to give the Einstein equations of GR in which the energy-momentum tensor is modified by corrections arising from the spin tensor \cite{review,Niko,Gas}.
These corrections are significant only at extremely high densities which are much higher than the nuclear density, on the order of $10^{45} \textrm{kg/m}^3$ \cite{Dir,torsion1}.
Below this density, the predictions of the ECSK theory are physically indistinguishable from the predictions of GR and reduce to them in vacuum, where torsion vanishes.

Fermions are described in relativistic quantum mechanics by the Dirac equation.
Since Dirac fields couple minimally to the torsion tensor, fermions are the source of torsion \cite{ScKi1,ScKi2,review,Niko}.
At macroscopic scales, these particles can be averaged as a spin fluid \cite{HHK,Gas,Dir,Kuc,torsion1}, which results from the multipole expansion of the conservation law for the spin tensor in the ECSK theory \cite{Niko,NSH}.
Even if the spin orientation of particles is random, the terms with the spin tensor in the field equations are quadratic and do not vanish in a spin fluid.
The Einstein-Cartan equations for a spin fluid are equivalent to the Einstein equations for an ideal fluid with the effective energy density $\tilde{\epsilon}$ and pressure $\tilde{p}$ given by \cite{Niko,HHK,Gas,Kuc,torsion1}
\begin{eqnarray}
\tilde{\epsilon}=\epsilon-\alpha n_\textrm{f}^2,\\
\tilde{p}=p-\alpha n_\textrm{f}^2,
\end{eqnarray}
where $n_\textrm{f}$ is the number density of fermions and $\alpha=\kappa(\hbar c)^2/32$.

The spin fluid in the early Universe is formed by an ultrarelativistic matter in kinetic equilibrium, for which $\epsilon=h_\star T^4$, $p=\epsilon/3$ and $n_\textrm{f}=h_{n\textrm{f}}T^3$, where $T$ is the temperature of the Universe, $h_\star=(\pi^2/30)(g_\textrm{b}+(7/8)g_\textrm{f})k_\textrm{B}^4/(\hbar c)^3$, and $h_{n\textrm{f}}=(\zeta(3)/\pi^2)(3/4)g_\textrm{f}k_\textrm{B}^3/(\hbar c)^3$.
For standard-model particles, $g_\textrm{b}=28$ and $g_\textrm{f}=90$.
If we assume that the Universe is closed, homogeneous, and isotropic, then it is described by the Friedmann-Lema\^{i}tre-Robertson-Walker (FLRW) metric.
In the presence of spin and torsion, the first Friedmann equation can be written as \cite{torsion2,torsion2014}
\begin{equation}
\frac{{\dot{a}}^2}{c^2}+k=\frac{1}{3}\kappa\tilde{\epsilon}a^2=\frac{1}{3}\kappa(h_\star T^4-\alpha h_{n\textrm{f}}^2 T^6)a^2,
\label{nikoeq1}
\end{equation}
where $a$ is the scale factor, $k=1$, and dot denotes the derivative with respect to the cosmic time.
The second Friedmann equation can be written as a continuity equation which in the presence of particle production becomes \cite{torsion2014}
\begin{equation}
\frac{\dot{a}}{a}+\frac{\dot{T}}{T}=\frac{cK}{3h_{n1}T^3},
\label{nikoeq2}
\end{equation}
where $h_{n1}=(\zeta(3)/\pi^2)g_{n1}k_\textrm{B}^3/(\hbar c)^3$ and $g_{n1}=9$.
A scalar $K$ has the dimension of m$^{-4}$ and is related to the square of the curvature tensor \cite{Par,Zel,ZeSt,prod}.
For a rigorous treatment, it should be derived from quantum field theory in the Riemann-Cartan spacetime.
Following \cite{torsion2014}, we assume that
\begin{equation}
K=\beta(\kappa\tilde{\epsilon})^2,
\label{rate}
\end{equation}
where $\beta$ is a dimensionless particle production coefficient.
Equations (\ref{nikoeq1}) and (\ref{nikoeq2}) describe the dynamics of the early Universe.

Since the negative term on the right-hand side of Eq. (\ref{nikoeq1}) scales with $T$ faster ($\sim T^6$) than the positive term ($\sim T^4$), $\dot{a}$ reaches zero and the universe undergoes a nonsingular bounce at a positive minimum scale factor.
The temperature at a bounce is the maximum temperature of the Universe.
This temperature is given by Eq. (\ref{nikoeq1}) with $\dot{a}=0$ in which we can omit $k$ (which is negligible relative to the other terms) \cite{torsion2,torsion2014}:
\begin{equation}
T_\textrm{max}=\Bigl(\frac{h_\star}{\alpha h_{n\textrm{f}}^2}\Bigr)^{1/2}.
\label{Tmax}
\end{equation}
For standard-model particles, it is equal to $1.1524 \times 10^{32}$ K.
Particle production should vanish at a bounce, otherwise the temperature at that instant could exceed $T_\textrm{max}$ which would contradict Eq. (\ref{nikoeq1}).
This condition justifies the choice of the scalar $K$ in Eq. (\ref{rate}).

The contraction of the Universe before the Big Bounce could correspond to gravitational collapse of matter inside a black hole existing in another universe \cite{torsion1,Pat,Smo}.
During the collapse of a dustlike medium in GR, each spatial point in the interior of a black hole locally evolves toward a singularity as an independent, spatially homogeneous and isotropic universe \cite{LL}.
Numerical analysis of generic gravitational collapse shows that spatial derivatives of the metric and curvature tensors can be neglected and each spatial point in the interior of a black hole locally evolves as a spatially homogeneous universe \cite{num}.
In GR, a singularity forms before the collapse has completed \cite{LL,dyn}.
The infalling matter makes the interior of the black hole a dynamical spacetime with large inhomogeneities and anisotropies.
During the collapse, multiple trapped null surfaces with a very complicated structure and at least two dynamical wormhole throats form in the interior \cite{dyn}.
In the ECSK theory, we expect that each spatial point evolves toward a state of an extremely high but finite density and curvature, avoiding a singularity \cite{torsion2014}.
At such a state, the local contraction ends, the matter undergoes a bounce, and the local expansion begins.
Quantum effects in the presence of an extremely strong gravitational field cause an intense particle production \cite{Par,Zel,ZeSt,prod}, which creates an enormous amount of mass without changing the total energy (matter plus gravitational field) in the black hole \cite{torsion4}.

We conjecture that eventually the wormholes merge into one wormhole and the outermost trapped surface becomes an event horizon.
Asymptotically, the throat of this wormhole and the event horizon coincide.
The entire interior of a black hole would then become a new, closed universe whose dynamics cannot be observed from the outside of the black hole because of an infinite redshift at its event horizon.
This universe can be thought of as a three-dimensional analogue of the two-dimensional surface of a sphere \cite{torsion1}.
We posit that particle production in such a universe would make it effectively homogeneous and isotropic \cite{torsion2014}.

After a bounce, the universe expands and its temperature decreases to the values at which $k$ cannot be neglected in Eq. (\ref{nikoeq1}).
Eventually, $\dot{a}$ reaches zero and the universe undergoes a crunch at a maximum scale factor.
The universe then contracts until it reaches another bounce, and expands again.
Because of particle production near a bounce, the scale factor at a given bounce is larger than the scale factor at the preceding bounce.
The scale factor at a given crunch is larger than the scale factor at the preceding crunch.
When the universe produces sufficient amounts of mass, it reaches the size at which the temperature decreases to the matter-radiation equality value for which the energy density of nonrelativistic matter exceeds the energy density of radiation.
The universe then expands according to the first Friedmann equation in which the energy density is dominated by nonrelativistic matter.
The universe has contracting and expanding phases until the scale factor reaches the size at which the energy density for the cosmological constant exceeds the energy density of nonrelativistic matter.
The universe then begins to accelerate and expands indefinitely.

The last bounce (if there are more than one) is the Big Bounce that corresponds to the Big Bang \cite{torsion2014}.
It is followed by the accelerating torsion-dominated era (inflation), which lasts for a finite time because $\alpha h_{n\textrm{f}}^2 T^6$ in Eq. (\ref{nikoeq1}) eventually becomes negligible relative to $h_\star T^4$.
The universe then enters the radiation-dominated era (decelerating) without needing reheating, which is followed by the matter-dominated era (decelerating) and cosmological-constant era (accelerating).
Accordingly, our Universe may have been formed in a black hole existing in another universe \cite{torsion1,torsion2014}.
We note that, although this scenario can naturally explain the origin of the Universe, Eqs. (\ref{nikoeq1}) and (\ref{nikoeq2}) can describe the dynamics of the early Universe even if we do not assume a black hole as its origin.


\section{Inflationary dynamics}
The dynamics of a universe inside a black hole with torsion is encapsulated in Eqs. (\ref{nikoeq1}) and (\ref{nikoeq2}).
Near a bounce, where we can neglect $k$, these equations and Eq. (\ref{rate}) give \cite{torsion2014}
\begin{equation}
\frac{\dot{a}}{a}\Bigl[1-\frac{3\beta}{c^3 h_{n1}T^3}\Bigl(\frac{\dot{a}}{a}\Bigr)^3\Bigr]=-\frac{\dot{T}}{T}.
\label{ineq}
\end{equation}
The signs of $\dot{a}$ and $\dot{T}$ must be opposite to avoid an indefinite increase of the scale factor.
Thus, during an expanding phase, the second term in the square bracket in Eq. (\ref{ineq}) should be less than 1.
This term has a maximum at $T=T_\textrm{max}/\sqrt{2}$.
Therefore, the value of the particle production coefficient must satisfy \cite{torsion2014}
\begin{equation}
\beta<\beta_\textrm{cr}=\frac{\sqrt{6}}{32}\frac{h_{n1}h_{n\textrm{f}}^3(\hbar c)^3}{h_\star^3}.
\end{equation}
For standard-model particles, $\beta_\textrm{cr}=1/929.0915$.

For $\beta=0$, the universe in a black hole is oscillatory with an infinite number of bounces and crunches (cycles) \cite{torsion2014}.
If $0<\beta<\beta_\textrm{cr}$, then the universe is cyclic with a finite number of cycles, after which it expands indefinitely.
As $\beta$ increases, the number of cycles decreases and the early accelerated expansion of the universe in each cycle is closer to exponential.
If $\beta$ is slightly less than $\beta_\textrm{cr}$, then the universe has only one bounce.
In this case, Eq. (\ref{ineq}) at $T=T_\textrm{max}/\sqrt{2}$ gives $\dot{T}\approx 0$ and $\dot{a}/a\approx$ const.
Accordingly, the universe has a finite period of a nearly exponential expansion at a nearly constant energy density \cite{torsion2014}
\begin{equation}
\tilde{\epsilon}\approx\frac{h_\star^3}{8\alpha^2 h_{n\textrm{f}}^4}.
\end{equation}
This period is a part of the torsion-dominated era.
If $\beta\ge\beta_\textrm{cr}$, then an exponential expansion of the universe would last indefinitely (eternal inflation).
The number of bounces before the universe reaches the matter-radiation equality temperature $T_\textrm{eq}=8820$ K \cite{WMAP} as a function of $\beta/\beta_{cr}$ is shown in Table~\ref{table1}.

\begin{table}
\begin{center}
\begin{small}
\begin{tabular}{ |c|c|}
\hline
$\beta/\beta_{cr}$  & Number of bounces \\
\hline
0.996 & 1 \\
0.984 & 2  \\
0.965 & 3 \\
0.914 & 5 \\
0.757 & 10 \\
\hline
\end{tabular}
\caption{The number of bounces before the universe reaches the matter-radiation equality temperature as a function of the ratio of the particle production coefficient to its critical value ($\beta/\beta_{cr}$).}
\label{table1}
\end{small}
\end{center}
\end{table}

Now, we find the time dependence of the scale factor and its time derivative by numerically solving Eqs. (\ref{nikoeq1}) and (\ref{nikoeq2}) with a given set of the initial conditions.
We need an initial condition for $a_0$ at the Big Bounce.
We estimate that the minimum value of $a_0$ would be on the order of the Cartan radius of an electron which is on the order of $10^{-27}$ m \cite{torsion2}.
We need a value of $\beta$ which is slightly smaller than $\beta_{cr}$.
Thus, we choose $\beta=1/929.25$ and $a_0=10^{-27}$ m.
The initial temperature $T_0$ should be slightly less than $T_\textrm{max}$ given by Eq. (\ref{Tmax}).
We choose $T_0=0.99 T_\textrm{max}$.
As long as $T_0$ is near $T_\textrm{max}$, the dynamics of the expansion is not sensitive to the exact value of $T_0$.
We shall show later that the dynamics of such a universe is also insensitive to the initial values $a_0$, as long as $\beta \lesssim \beta_\textrm{cr}$. 

\begin{figure}
\centering
\includegraphics[width=0.5\textwidth]{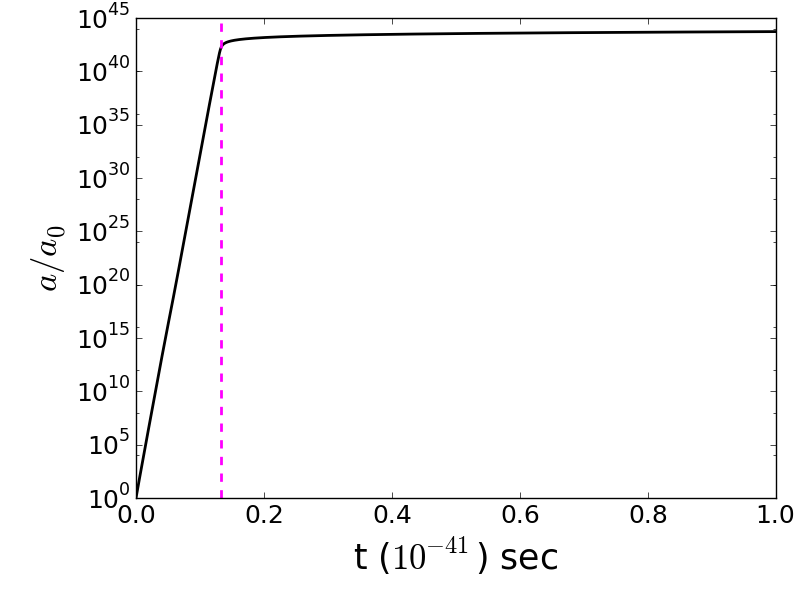}
\caption{The ratio of the comoving scale factor $a(t)$ to its initial value as a function of time.
The dashed magenta line at $t\sim 1.33 \times 10^{-42}$ seconds represents the transition from acceleration to deceleration.
We obtain about 60 $e$-folds.
The time $t=0$ is set at the Big Bounce.}
\label{fig1}
\end{figure}

\begin{figure}
\centering
\includegraphics[width=0.5\textwidth]{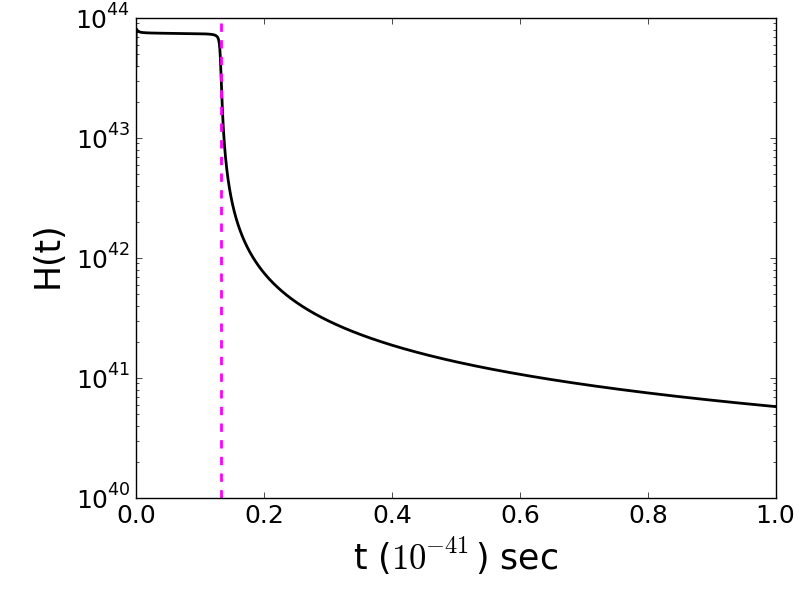}
\caption{The Hubble parameter as a function of time.
The dashed magenta line is at the same location as in Fig.~\ref{fig1}.
The Hubble parameter is roughly constant during the inflationary phase.} 
\label{fig2}
\end{figure}

\begin{figure}
\centering
\includegraphics[width=0.5\textwidth]{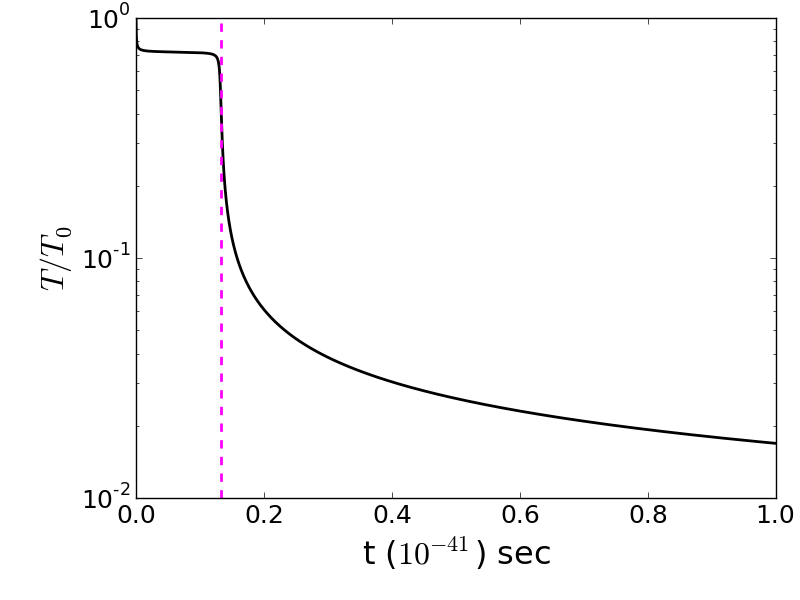}
\caption{The ratio of the temperature of the universe to its initial value as a function of time.
The dashed line is at the same location as in Fig.~\ref{fig1}.}
\label{fig3}
\end{figure}

With these initial conditions, the numerical solutions of Eqs. (\ref{nikoeq1}) and (\ref{nikoeq2}) for $a(t)$, $H(t)$, and $T(t)$ starting from the last bounce are shown as functions of time in Fig.~\ref{fig1},~\ref{fig2}, and ~\ref{fig3}, respectively.
As we can see from Fig.~\ref{fig1}, after the Big Bounce (shown by $t=0$ in these figures), the comoving scale factor expands exponentially and the universe is in an accelerating phase until $t=1.33 \times 10^{-42}$ seconds.
The total number of $e$-folds with these initial conditions and particle production coefficient  is about 60 from the start of the inflationary phase. 
The Hubble parameter is also roughly constant until $t=1.33 \times 10^{-42}$ seconds.
The comoving Hubble radius also decreases during this period.
Therefore, the dynamics of the scale factor immediately after the Big Bounce provides a realistic model of inflation.

\section{Reconstruction of a scalar field potential}
In order to compare our inflationary model with the observational data such as the tensor-to-scalar ratio,  spectral index of density perturbations, and its running, one needs to consider the quantized perturbations of the torsion field in a background FLRW universe.
This approach has been studied for a single Dirac field with torsion~\cite{Watanabe}.
For our model, we shall address this in a future publication.
Here, we find a scalar field potential which gives the same dynamics of the scale factor as the numerical solution for $a(t)$ computed in the preceding section.
This way, we can use the standard method of calculating quantum fluctuations of the scalar field and background spacetime to determine the scalar and tensor perturbations~\cite{Mukhanov} and compare them with observations.

The procedure for finding a scalar field corresponding to a given dynamics of the scale factor was studied by Ellis and Madsen~\cite{EllisMadsen}.
Given any functional form of $a(t)$  on a background FLRW spacetime, which results in a consistent cosmological evolution, one can always construct a potential $V(\phi)$  corresponding to a scalar field $\phi(t)$ (inflaton)~\cite{EllisMadsen,Paddy}.
To calculate an inflaton potential, we reconstruct a scalar field potential from the evolution of the Hubble parameter in our model and calculate its slow-roll parameters.
We then calculate the tensor-to-scalar ratio $r$, scalar spectral index of density perturbations $n_s$ and its running $\alpha_s$ using these calculated values of the slow-roll parameters.
The equations from~\cite{EllisMadsen}  (their equations 17-18)  which are used to reconstruct a scalar field potential $V(\phi)$ from the expansion history of the universe on a FLRW background are:
\begin{equation}
V(t) = (m_\textrm{P}^2/8\pi)(\dot{H} +3 H^2+2c^2/a^2) ,
\label{V}
\end{equation}
\begin{equation}                         
\dot{\phi}^2 (t) = (m_\textrm{P}^2/4\pi) (-\dot{H} +c^2/a^2).
\label{phi}
\end{equation}
Here, $V(t)$ is the potential, $\phi(t)$ is the scalar field, and $m_\textrm{P}$ is the Planck mass \cite{Kinney}.
The third/second term in the right-hand sides of Eqs.~\ref{V} and~\ref{phi} represent the curvature terms.
These terms are usually omitted in models of inflation, since it is assumed that the process of inflation  drives the universe towards flatness.

\begin{figure}
\centering
\includegraphics[width=0.5\textwidth]{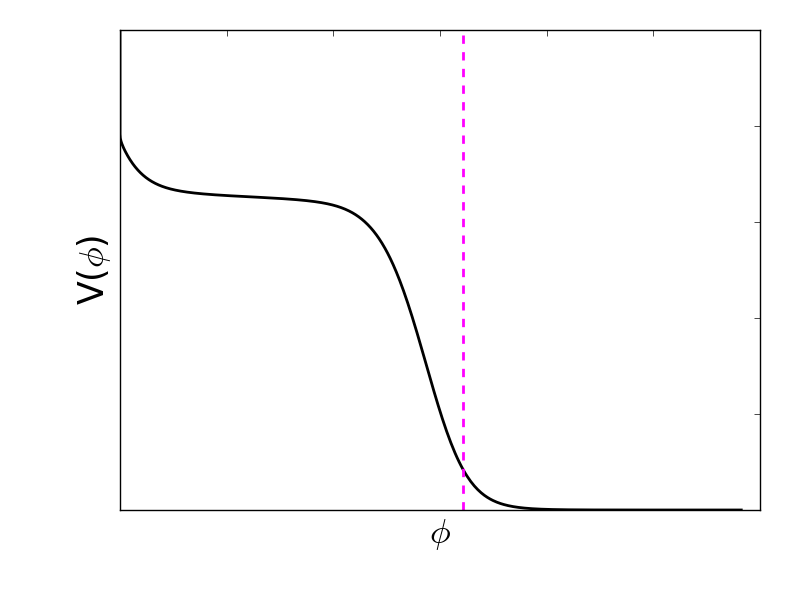}
\caption{The reconstruction of single-field scalar potential $V(\phi)$ from the inflationary dynamics depicted in Fig.~\ref{fig1},~\ref{fig2}, and~\ref{fig3}. 
The vertical line indicates the end of inflation in our model when the universe transitions to a decelerating phase.
Since we are only interested in the shape of the potential, we do not show units on the axes.}
\label{fig4}
\end{figure}

From the non-parametric expressions for $a(t)$, $H(t)$, and $\dot{H(t)}$, one can use Eqs.~\ref{V} and~\ref{phi} to obtain expressions for $V(t)$ and $\dot{\phi}^2(t)$.
The integration of $\dot{\phi}$ gives $\phi(t)$ which, combined with $V(t)$, gives an expression for $V(\phi)$ which gives the same dynamics of the scale factor as that found in the preceding section.
This inflaton potential is shown in Fig.~\ref{fig4}.
At early times we obtain a nearly flat potential which satisfies $V'(\phi) \ll V(\phi)$, which is then followed by a scalar field rolling down the potential.
The universe transitions from an acceleration to deceleration near the minimum of the potential.
In scalar field models of inflation, the height of the potential usually corresponds to the vacuum energy density and its width to the change in value of the scalar field during inflation.
The energy scale is usually associated with particle physics phenomenology.
However, since we want to find nondimensional quantities that characterize the CMB, we are only interested in the shape of $V(\phi)$ and not its absolute scale.
Hence, we ignore the absolute scale of the potential.

Using $a(t)$ and $V(\phi)$, we now calculate the slow-roll parameters $\epsilon$, $\eta$, $\eta_v$, and $\xi$ as functions of the number of $e$-folds before end of inflation, in order to make observable predictions for the spectrum of tensor and scalar perturbations. 
We use the expressions for $\epsilon$ and $\eta$ from~\cite{Kinney}, whereas $\eta_v$, and $\xi$ are obtained from~\cite{Saridakis}:
\begin{eqnarray}
\epsilon & = & 1- \frac{\ddot{a} a}{\dot{a}^2}, \label{epsilon} \\  
\eta & = & -\frac{\ddot{\phi}}{H\dot{\phi}}, \label{eta} \\
\eta_v & = & \frac{m_\textrm{P}^2}{8\pi V} \frac{d^2V}{d\phi^2}, \label{eta_v} \\
\xi & = & \frac{m_\textrm{P}^4}{64 \pi^2 V^2} \frac{dV}{d\phi} \frac{d^3 V}{d\phi^3}. \label{xi} 
\end{eqnarray}
Note that both $\eta$ and $\eta_v$ are the second slow-roll parameters, where $\eta$ is denoted as the Hubble slow-roll parameter and $\eta_v$ as the potential slow-roll parameter.
The tensor-to-scalar ratio $r$, scalar spectral index of curvature perturbations $n_s$, and the running of the spectral index, $\alpha_s \equiv d\ln n_s/d\ln k$, where $k$ is the comoving wavenumber, can be calculated from the above parameters using the slow-roll approximation~\cite{Kinney,Saridakis}:
\begin{eqnarray}
n_s & \approx & 1 - 4 \epsilon + 2 \eta, \label{ns}\\
r & \approx  & 16\epsilon, \label{r} \\
\alpha_s  & \approx &  16\epsilon\eta_v -24\epsilon^2 -2\xi. \label{alphas}  
\end{eqnarray}
To compare these quantities with observations, one calculates the slow-roll parameters for the value of $N$ (number of $e$-folds before end of inflation) when the present Hubble scale crossed outside the horizon during inflation.
For most inflationary models, $N$ is about 50-60~\cite{Planck2013,Planck2015}, depending on the dynamics of the reheating era.
However, the lower limit on the horizon crossing scale can be as low as $N=18$~\cite{Liddle}.
We shall calculate more precise bounds on this scale for our model in a forthcoming work.

\begin{figure}
\centering
\includegraphics[width=0.5\textwidth]{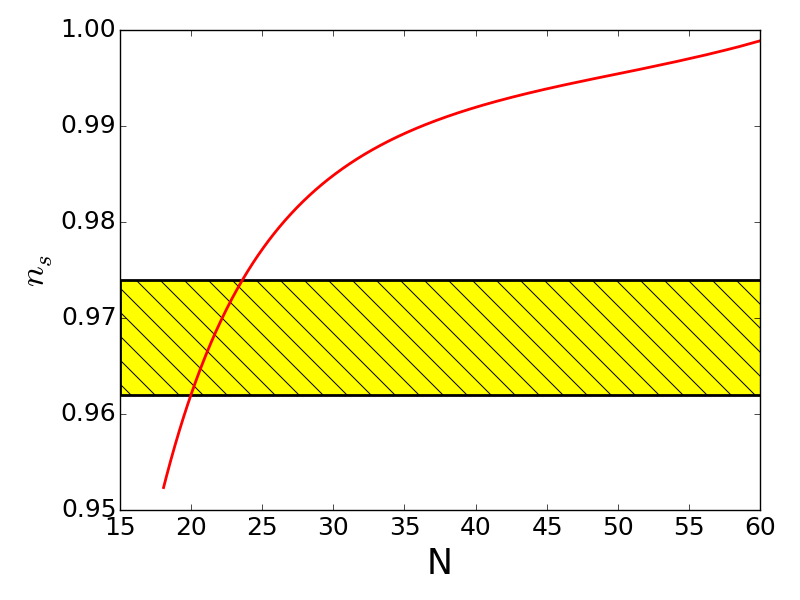}
\caption{The scalar spectral index  $n_s$ as a function of the number of $e$-folds before the end of inflation ($N$), evaluated using Eq.~\ref{ns}.
The hatched regions show the $1\sigma$ estimates for $n_s$ from the Planck 2015 TT + lowP + lensing data release~\cite{Planck2015cos} which is $n_s=0.968 \pm 0.006$.}
\label{fig5}
\end{figure}

\begin{figure}
\centering
\includegraphics[width=0.5\textwidth]{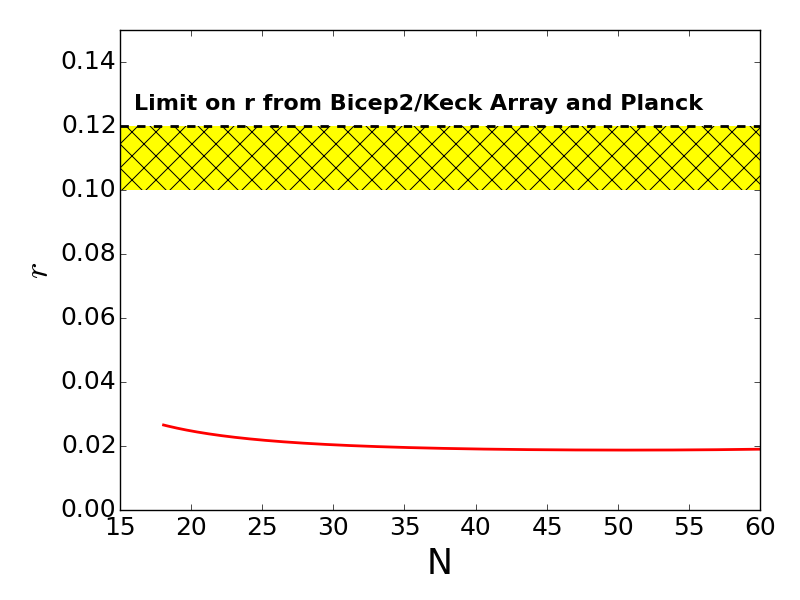}
\caption{The solid red curve shows the tensor-to-scalar ratio $r$ as a function of the number of $e$-folds before the end of inflation ($N$), evaluated using Eq.~\ref{r}.
The dashed black line shows the upper limit of $r_{0.05}<0.12$ at 95\% confidence level (at the pivot scale of 0.05 $\textrm{Mpc}^{-1}$), which is obtained from a joint analysis of BICEP2/Keck Array and the 2015 Planck observations~\cite{bicepkeck}.}

\label{fig6}
\end{figure}

\begin{figure}
\centering
\includegraphics[width=0.5\textwidth]{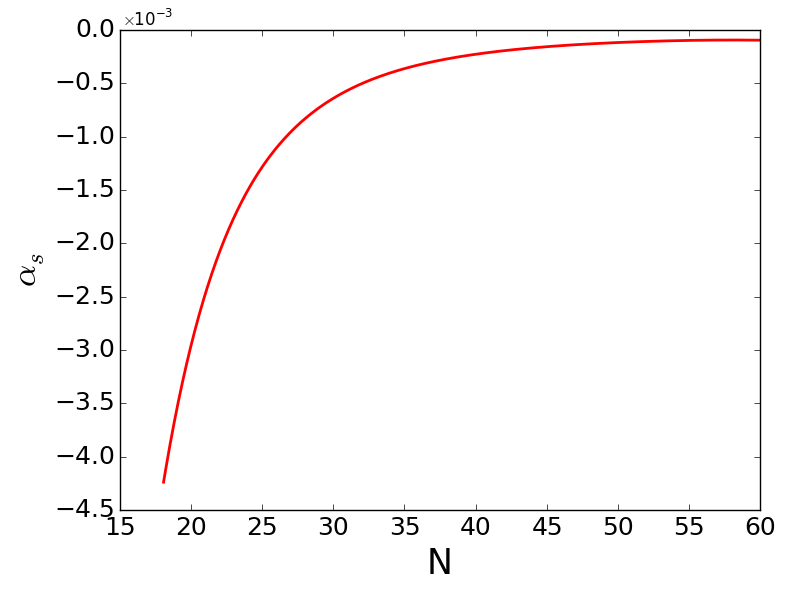}
\caption{The solid red curve shows the running of the spectral index ($\alpha_s$) as a function of the number of $e$-folds before the end of inflation ($N$), evaluated using Eq.~\ref{alphas}.
The Planck 2015 results are consistent with no running~\cite{Planck2015cos}.}
\label{fig7}
\end{figure}

For a conservative estimate, we evaluate the slow-roll parameters for values of N between 18 and 60, and use these parameters to obtain $n_s$, $r$, and $\alpha_s$ from Eqs~(\ref{ns}), ~(\ref{r}), and (\ref{alphas}) respectively.
The scalar spectral index $n_s$ obtained using Eq.~\ref{ns} is shown in Fig.~\ref{fig5}.
We always can observe a red tilt.
For values of $N$ between about 20 and 25, we obtain $n_s \approx 0.96$, which is consistent with the best-fit estimate of $n_s=0.968 \pm 0.006$ from Planck 2015 ~\cite{Planck2015cos}.
However, if the horizon crossing scale in our model occurs for the same values of $N$ as in most other inflationary models (between 50-60), we would obtain a value of $n_s \approx 0.99$, which on face value is in tension with the Planck 2015 best-fit estimate at about 6$\sigma$.
However, if we assume an extra relativistic degree of freedom, then the best fit estimate of $n_s$ from Planck 2015 is closer to our estimate of $\approx$ 0.99~\cite{Planck2015cos}.
The tensor-to-scalar ratio $r$ obtained using Eq.~\ref{r} is shown in Fig.~\ref{fig6}.
Our estimated value of $r$ for all permissible values of $N$ is about an order of magnitude smaller than the 95\% confidence upper limit on $r_{0.05}<0.12$, where $r_{0.05}$ is evaluated at the pivot scale of 0.05 $\textrm{Mpc}^{-1}$.
This limit is obtained from a joint analysis of Planck 2015 and BICEP2/KECK Array~\cite{bicepkeck}.
The running of the spectral index $\alpha_s$ is shown in Fig.~\ref{fig7}.
Our estimated value for the  running is about $\mathcal{O}(10^{-3})$, which  is consistent with the no running of the spectral index found from Planck 2015 observations~\cite{Planck2015cos}.
We also note that our values of $r$ and $\alpha_s$ are of the same order of magnitude as in Starobinsky's model of inflation \cite{Starobinsky}.

{\vskip 0.2cm}


\section{Sensitivity to initial conditions}
\begin{figure}
\centering
\includegraphics[width=0.5\textwidth]{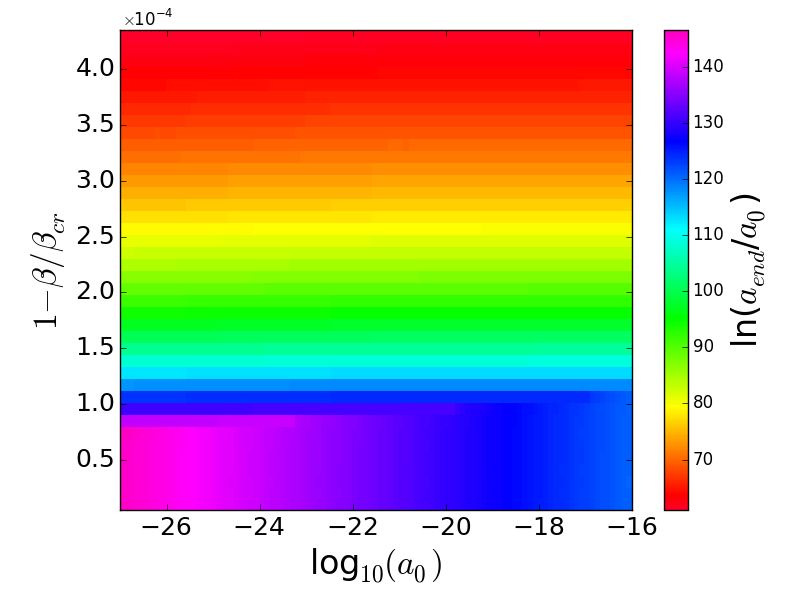}
\caption{The total number of expansion $e$-folds since the beginning of an inflationary phase as a function of the initial scale factor $a_0$ (in meters) and one minus the ratio of the particle production coefficient to its critical value $\beta/\beta_{cr}$.}
\label{evar}
\end{figure}

\begin{figure}
\centering
\includegraphics[width=0.5\textwidth]{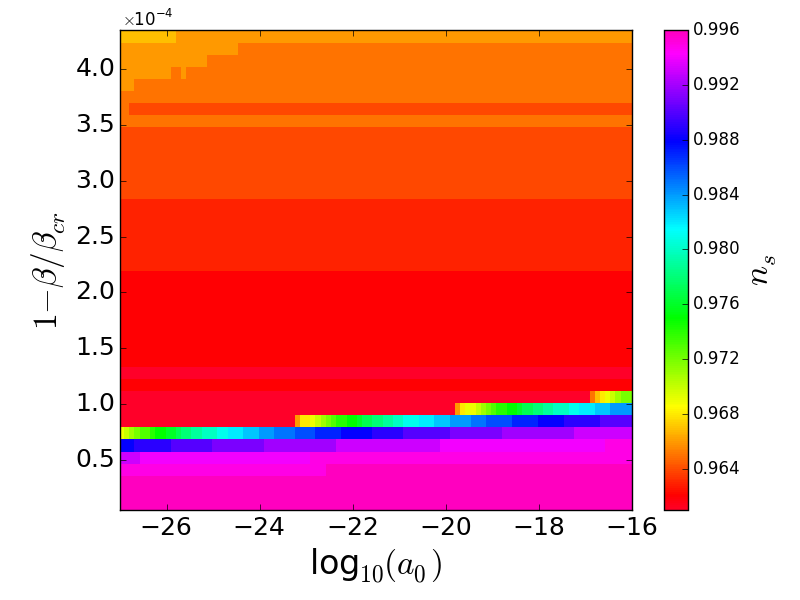}
\caption{The scalar spectral index ($n_s$) evaluated 20 $e$-folds before the end of inflation as a function of the initial scale factor $a_0$ (in meters) and one minus the ratio of the particle production coefficient to its critical value $\beta/\beta_{cr}$.}
\label{nsvar}
\end{figure}

\begin{figure}
\centering
\includegraphics[width=0.5\textwidth]{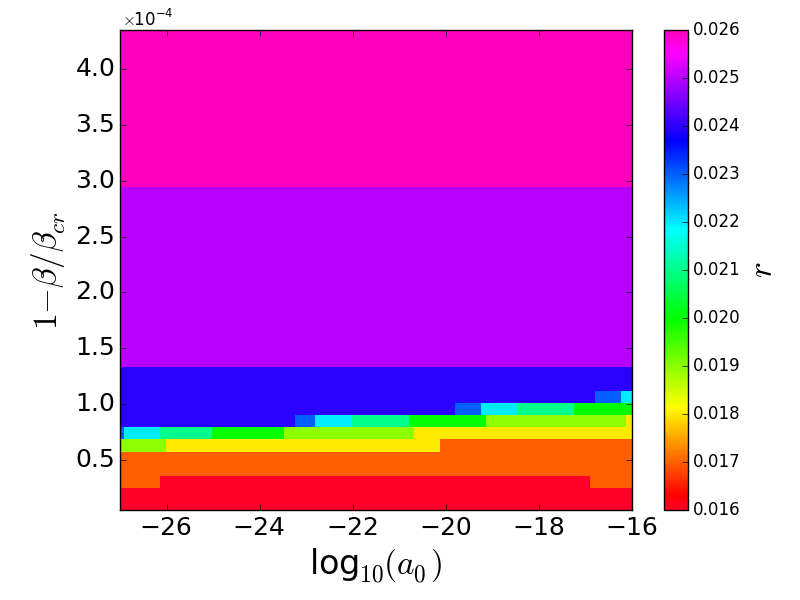}
\caption{The tensor-to-scalar ratio ($r$) evaluated 20 $e$-folds before the end of inflation as a function of the initial scale factor $a_0$ (in meters) and one minus the ratio of the particle production coefficient to its critical value $\beta/\beta_{cr}$.}
\label{rvar}
\end{figure}

\begin{figure}
\centering
\includegraphics[width=0.5\textwidth]{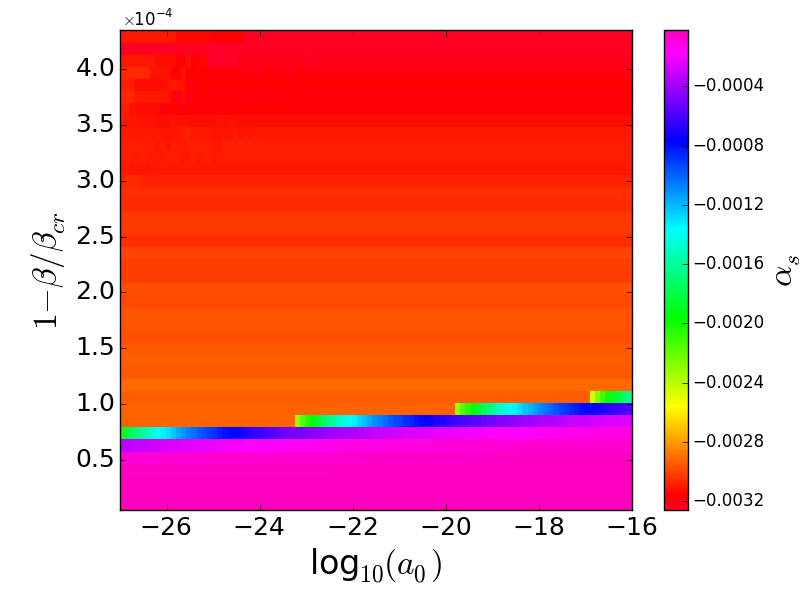}
\caption{The running of the spectral index ($\alpha_s$) evaluated 20 $e$-folds before the end of inflation as a function of the initial scale factor $a_0$ (in meters) and one minus the ratio of the particle production coefficient to its critical value $\beta/\beta_{cr}$.}
\label{alphavar}
\end{figure}

After calculating the expansion dynamics and the inflationary observables for one set of values of $a_0$ and $\beta$, we now examine the sensitivity of the dynamics to initial conditions by varying $a_0$ and $\beta$.
We do not vary $T_0$, since it must be close to $T_\textrm{max}$ in order to obtain inflation-like behavior.
The total number of $e$-folds from the start of an inflationary phase as a function of $a_0$ and $\beta/\beta_{cr}$ is shown in Fig.~\ref{evar}.
The total number of $e$-folds is insensitive to initial values of $a_0$ and only depends on $\beta$.
We now do the same exercise for $r$, $n_s$, and $\alpha_s$, where we evaluate these quantities 20 $e$-folds before the end of inflation for the same range of $a_0$ and $\beta$.
These variations are shown in Figs.~\ref{nsvar},~\ref{rvar}, and ~\ref{alphavar} respectively.
We always find a red tilt in the spectrum of scalar perturbations for different values of $a_0$ and $\beta$. The value of the tensor-to-scalar ratio is approximately constant and is between 0.01 and 0.03. The running of the spectral index is always less than 0 and its absolute value is $\mathcal{O}(10^{-3}-10^{-4})$
Therefore, the values of $n_s$,$r$, and $\alpha_s$ do not change significantly with $a_0$ and are only sensitive to $\beta$.
The total number of $e$-folds also effectively depends only on $\beta$.
Therefore, our inflationary model does not rely on the initial scale factor $a_0$ but depends on the particle production coefficient $\beta$.

\section{Other models of inflation with spin}
The first model of inflation based on the ECSK theory was by Gasperini~\cite{Gas}, who also has considered the dynamics of a spin fluid in the FLRW spacetime.
In that model, a spin fluid with $w=-1/3+\delta$, where $w=p/\epsilon$ and $0<\delta\ll 1$ is a number extremely close to zero, can characterize a physically viable inflationary scenario with three distinct phases of acceleration which can give about 70 $e$-folds.
The origin of such a fine tuning would be, however, difficult to explain.
Obukhov~\cite{Obukhov} has considered a GR fluid with the spin-spin interactions characterized by an effective potential, which was chosen to be similar to that used in vector field models of inflation.
That model can give an inflationary phase which lasts for about $10^{-43}$ s, but does not specify the origin of such a potential.
Watanabe~\cite{Watanabe} has considered a single Dirac spinor field in the ECSK theory, characterized by a generic potential which depends on the scalar constructed from the spinors.
That model can give a viable inflationary phase with an almost scale invariant spectrum of perturbations of the Dirac field on the FLRW background and a de Sitter expansion phase with more than 60 $e$-folds if such a potential is appropriately chosen.
Most recently, Choudhury {\it et al.}~\cite{Choudhury} have considered an effective potential for a scalar field associated with torsion and its quantum corrections.
That model gives an inflationary phase with $n_s \approx 0.96$, $r \approx 0.056$, and $\alpha_s \simeq -10^{-4}$.
There also exist models of inflation based on fermion fields with a symmetric affine connection and other theories with torsion such as teleparallel gravity or f(T) gravity.
In those theories, however, the orbital angular momentum is conserved, which contradicts the exchange between the orbital and spin parts of the angular momentum, that happens for electrons in atoms~\cite{req}.

The model presented in this paper is similar to that in \cite{Gas,torsion1} with the addition of the temperature dependences of the terms in the energy-momentum tensor \cite{torsion2}.
The spin-spin corrections to the energy-momentum tensor arising from the minimal coupling between spin and torsion in the ECSK theory give the dynamics of the scale factor that avoids the singularity and solves the flatness and horizon problems without constraining the equation of state of matter.
Adding quantum particle production can explain the origin of matter in the Universe and generate an inflationary phase without constraining a potential driving inflation.
The only relevant parameter in this scenario is the production coefficient $\beta$ which ultimately should be derived from quantum gravity.


\section{Conclusions}
We studied the dynamics of a universe formed in a black hole, using the Einstein-Cartan-Sciama-Kibble theory of gravity which extends general relativity by including the spin of matter and torsion of spacetime.
We solved numerically the equations describing such a universe in the torsion-dominated era which follows the last bounce.
We demonstrated that for values of the particle production coefficient below a critical value, the dynamics of the scale factor is similar to that in typical models of cosmic inflation.
We obtain about 60-150 $e$-folds, depending on the value of the production coefficient.
From the calculated expansion of the scale factor, we reconstructed a dynamically equivalent scalar field and the corresponding inflaton potential, and showed that the shape of such a potential is similar to those in most generic models of inflation.
We also demonstrated that the dynamics of the universe is insensitive to the initial values of the scale factor and effectively depends on the particle production coefficient only.

We calculated the slow-roll parameters corresponding to the reconstructed scalar field.
Using the slow-roll approximations, we found the tensor-to-scalar ratio, scalar spectral index of curvature perturbations, and its running as functions of the number of $e$-folds before the end of inflation.
Our values of the tensor-to-scalar ratio are about ten times smaller than the 95 \% confidence level limits obtained from the joint analysis of Bicep-2/Keck and Planck data.
The values of the scalar spectral index agree with the Planck 2015 results when evaluated about 20 $e$-folds before the end of inflation for a particular range of the particle production coefficient.
We also find that the  running of the scalar spectral index is negligible. 
Therefore, a universe in a black hole with spin, torsion, and particle production provides a simple and natural mechanism for inflation which does not require hypothetical fields and is consistent with the Planck 2015 observations.
Other observables such as non-gaussianities of the CMB radiation will be considered in future work.

\section{Acknowledgments}

We would like to thank I-Non Chiu and  Valeria Pettorino for helpful correspondence.


\end{document}